\documentclass[twocolumn,showpacs,preprintnumbers,amsmath,amssymb]{revtex4}
\usepackage{graphicx}% Include figure files
\usepackage{dcolumn}% Align table columns on decimal point
\usepackage{bm}% bold math
\usepackage{booktabs}
\usepackage{multirow}
\usepackage{amssymb}
\begin{document}

%\preprint{submit to Appl. Phys. Lett.}

\title{Large enhancement in the generation efficiency of pure spin currents in Ge \\ using Heusler-compound Co$_{2}$FeSi electrodes}% Force line breaks with \\

\author{K. Kasahara,$^{1}$ Y. Fujita,$^{1}$ S. Yamada,$^{1}$ K. Sawano,$^{2}$ M. Miyao,$^{1}$ and K. Hamaya$^{1}$\footnote{E-mail: hamaya@ed.kyushu-u.ac.jp}}

\affiliation{$^{1}$Department of Electronics, Kyushu University, 744 Motooka, Fukuoka 819-0395, Japan}%
\affiliation{$^{2}$Advanced Research Laboratories, Tokyo City University, 8-15-1 Todoroki, Tokyo 158-0082, Japan}

%\affiliation{
%%Second institution and/or address\\

\date{\today}% It is always \today, today,
             
\begin{abstract}
We show nonlocal spin transport in {\it n}-Ge based lateral spin-valve (LSV) devices with highly ordered Co$_{2}$FeSi/{\it n}$^{+}$-Ge Schottky tunnel contacts. Clear spin-valve signals and Hanle-effect curves are demonstrated at low temperatures, indicating generation, manipulation, and detection of pure spin currents in {\it n}-Ge. The obtained spin generation efficiency of $\sim$0.12 is about two orders of magnitude larger than that for a device with Fe/MgO tunnel-barrier contacts reported previously. Taking the spin related behavior with temperature evolution into account, we infer that it is necessary to simultaneously demonstrate the high spin generation efficiency and improve the quality of the transport channel for achieving Ge based spintronic devices.
\end{abstract}
%\pacs{72.25.Dc, 72.25.Hg, 72.25.Mk}% PACS, the Physics and Astronomy
%\keywords{Suggested keywords}%Use showkeys class option if keyword display desired
                    
\maketitle

%\section{INTRODUCTION}
Semiconductor based spintronic devices have been proposed in a scalable solid state framework.\cite{Wolf,Datta,Dery1,Igor} For the compatibility with the existing electronic devices on semiconductor platforms, all electrical means of generation, transport and detection of spin polarized carriers through semiconductor channels such as GaAs,\cite{Lou}  Si,\cite{Appelbaum} Ge,\cite{Zhou} and so forth, are important technologies. To realize low power consumption for actual applications, the use of Schottky-tunnel contacts is more effective than that of insulating tunnel-barrier contacts because of low parasitic resistance in nano-scaled devices.\cite{HamayaJJAP,HamayaJAP,Lou,Ramsteiner} 

Recently, since high electron and hole mobilities are required as next-generation channel materials, Ge based spintronics compatible with Si-LSIs has also been explored by many groups.\cite{Zhou,Jeon,Saito,Jain,Kasahara2,Koike} So far, by using nonlocal spin-valve (NLSV) measurements in Ge based laterally configured devices, generation and detection of pure spin currents were observed up to $\sim$ 200 K.\cite{Zhou,Hamaya2,Chang} Unfortunately, above 100 K, there has not been spin manipulation via Hanle-type spin precession in nonlocal four-terminal geometries. Despite the devices with Fe/MgO tunnel-barrier contacts,\cite{Zhou,Chang} the Hanle-effect curves in nonlocal four-terminal geometries were able to be seen only at less than 50 K. In general, even at low temperatures, the spin lifetime of the {\it n}-Ge channels can be estimated frequently to be less than 1 ns,\cite{Zhou,Chang} which is significantly shorter than that of {\it n}-GaAs\cite{Lou} and {\it n}$^{+}$-Si.\cite{Shiraishi} The short spin lifetime can lead to short spin diffusion length in the channel region. This means that, in lateral device structures, the demonstration of the pure spin current transport in {\it n}-Ge is more difficult than that in {\it n}-GaAs and {\it n}$^{+}$-Si experimentally. 

Until now, highly ordered Co$_{2}$FeSi$_{1-x}$Al$_{x}$ Heusler compounds have been explored as the spin injector and detector to improve the spin generation efficiency in the GaAs channels\cite{Bruski,Tezuka1} although {\it n}-GaAs channels can not be compatible with Si-LSIs, where Co$_{2}$FeSi$_{1-x}$Al$_{x}$ compounds have relatively large spin polarization like a half-metallic material.\cite{Inomata,Sukegawa,Kimura1,Hamaya1} Since we have established the growth technique for highly ordered Co$_{2}$FeSi ($x=0$) on Ge by using low-temperature molecular beam epitaxy,\cite{Kasahara1} we can combine the Co$_{2}$FeSi electrodes with spin injection/detection techniques reported so far.\cite{Kasahara2,Hamaya2} If large spin accumulation is generated in the {\it n}-Ge channel by using the highly ordered Co$_{2}$FeSi, the lateral transport with the spin manipulation, i.e., Hanle-effect curve, can be observed at higher temperatures. 

In this study, using $L$2$_\text{1}$-ordered Co$_{2}$FeSi/{\it n}$^{+}$-Ge Schottky-tunnel contacts, we demonstrate lateral transport of pure spin currents detected by four-terminal nonlocal Hanle effects in {\it n}-Ge up to 225 K. The spin generation efficiency is markedly enhanced, which is about two orders of magnitude larger than that of the previous report by Zhou {\it et al}.\cite{Zhou} The use of Co based Heusler compounds will be a candidate for metallic source and drain in Ge based spintronic applications. We also discuss the spin related behavior with temperature evolution. 
\begin{figure}
\begin{center}
\includegraphics[width=7cm]{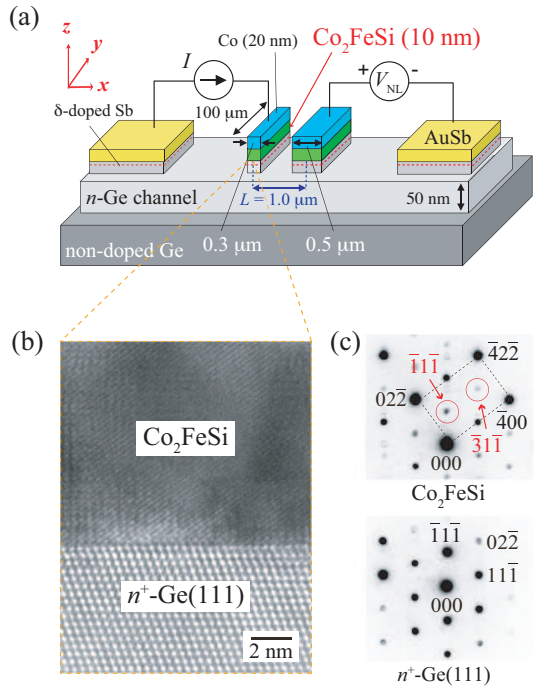}
\caption{(Color online) (a) Schematic diagram of a lateral four-terminal device with Co$_{2}$FeSi/{\it n}$^{+}$-Ge contacts. (b) A cross-sectional TEM image of the Co$_{2}$FeSi/{\it n}$^{+}$-Ge interface fabricated by a room-temperature MBE technique. (c) Nanobeam electron diffraction patterns for each Co$_{2}$FeSi and {\it n}$^{+}$-Ge. The zone axis is parallel to the [011] direction.}
\end{center}
\end{figure}

As schematically shown in Fig. 1(a), we fabricated {\it n}-Ge based lateral spin-valve devices (LSVs) with Co$_{2}$FeSi/{\it n}$^{+}$-Ge Schottky tunnel contacts. The followings are fabrication processes. First, we formed a phosphorous (P)-doped {\it n}-Ge(111) channel (P$^{+}$ $\sim$ 10$^{18}$ cm$^{-3}$) with a thickness of $\sim$50 nm on non-doped Ge(111) substrates ($\rho$ = $\sim$40 $\Omega$cm) by using an ion implantation technique and post annealing at 700 $^{\circ}$C. Then, an {\it n}$^{+}$-Ge(111) layer consisting of Sb $\delta$-doped layer and a 5-nm-thick Ge epitaxial layer was grown by molecular beam epitaxy (MBE) at 400 $^{\circ}$C,\cite{Sawano} where the doping density of  Sb was 2 $\times$ 10$^{14}$ cm$^{-2}$. After the fabrication of the {{\it n}$^{+}$-Ge(111) layer, a 10-nm-thick Co$_{2}$FeSi epitaxial layer was grown on top of the {\it n}$^{+}$-Ge(111) layer by room-temperature MBE.\cite{Yamada} In Fig. 1(b), we show a high-resolution cross-sectional transmission electron micrograph (TEM) of the formed Co$_{2}$FeSi/{\it n}$^{+}$-Ge interface. The heterojunction is atomically flat, leading to the reduction in the presence of interface states.\cite{Kasahara,Yamane} We also observe $<$111$>$ and $<$113$>$ superlattice reflections in the nanobeam electron diffraction patterns of the Co$_{2}$FeSi layer [Fig. 1(c)], resulting from the presence of $L$2$_\text{1}$-ordered structures. Thus, the Co$_{2}$FeSi Heusler-compound electrodes were high quality as shown in our  previous work.\cite{Kasahara1} In Fig. 1(c), the quality of the {\it n}$^{+}$-Ge layer is guaranteed. In order to align the magnetic moments in the in-plane direction for the Hanle-effect measurements, a polycrystalline Co layer with a thickness of $\sim$20 nm was deposited on the Co$_{2}$FeSi layer by using electron beam evaporation. 

Conventional processes with electron-beam lithography, Ar$^{+}$ ion milling, and reactive ion etching were used to fabricate four-terminal LSVs. The size of each contact is presented in Fig. 1(a) (0.3 $\times$ 100 $\mu$m$^{2}$ and 0.5 $\times$ 100 $\mu$m$^{2}$), and the center-to-center distance ($L$) between the Co$_{2}$FeSi/{\it n}$^{+}$-Ge contacts was 1.0 $\mu$m. The current-voltage ($I$-$V$) characteristics of the Co$_{2}$FeSi/{\it n}$^{+}$-Ge junctions showed almost no rectifying behavior at room temperature, indicating tunneling conduction of electrons through the Co$_{2}$FeSi/{\it n}$^{+}$-Ge interface. Although we confirmed no change in the forward-bias current with temperature variation, small decreases in the reverse-bias current were observed with decreasing temperature. Nearly same features have already been seen in our previous work for the {{\it n}$^{+}$-Ge(111) layer formed by Sb $\delta$-doping.\cite{Hamaya2} In order to avoid the change in the interface resistance with changing temperature, we concentrate on measurements in the forward-bias conditions for the Co$_{2}$FeSi/{\it n}$^{+}$-Ge contacts. In this study, the resistance area product ($RA$) of the Co$_{2}$FeSi/{\it n}$^{+}$-Ge interface was nearly constant of $\sim$ 10$^{4}$ $\Omega$$\mu$m$^{2}$. In order to precisely understand spin-related phenomena, we also fabricated a micro fabricated Hall-bar device with the same {\it n}-Ge channel and AuSb ohmic contacts. The electron carrier density ($n$) of the fabricated channel was experimentally estimated from electrical Hall-effect measurements. The estimated values were {\it n} $\sim$ 2.5 $\times$ 10$^{18}$ cm$^{-3}$ at 300 K and {\it n} $\sim$ 3.3 $\times$ 10$^{17}$ cm$^{-3}$ at 150 K, respectively. As a result, the electron mobility ($\mu_{e}$) of this channel is $\sim$ 286 cm$^{2}$V$^{-1}$s$^{-1}$ at 300 K and $\sim$ 665 cm$^{2}$V$^{-1}$s$^{-1}$ at 150 K. 
\begin{figure}
\begin{center}
\includegraphics[width=7cm]{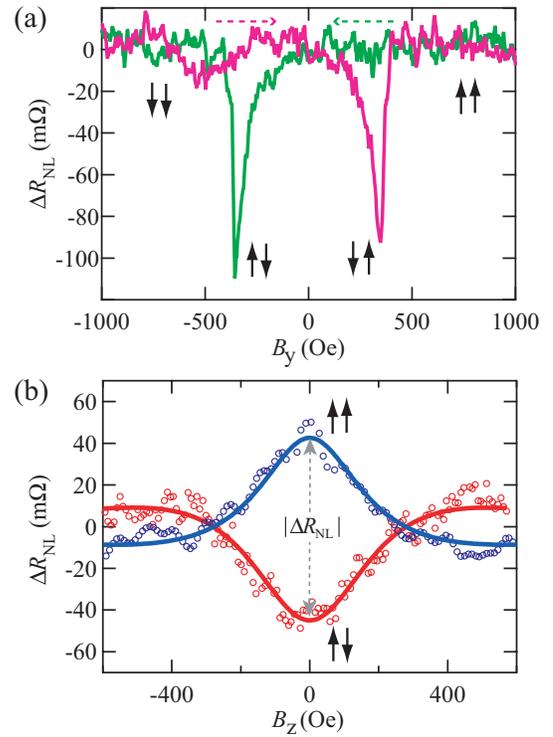}
\caption{(Color online) (a) Nonlocal magnetoresistance curve at 150 K. (b) Hanle-effect curves at 150 K for parallel and anti-parallel magnetization configuration. The $|\Delta R_{\rm NL}|$ value is defined from the Hanle curves.}
\end{center}
\end{figure}

Figure 2(a) shows a nonlocal magnetoresistance ($\Delta R_{\rm NL} =$$\frac{\Delta V_{\rm NL}}{I}$) measured at $I =$ +1.0 mA at 150 K, where the positive sign of $I$ ($I >$ 0) means that the electrons are extracted from the Ge channel into Co$_{2}$FeSi, i.e., spin extraction condition, through the Schottky tunnel barrier. By applying in-plane magnetic fields ($B_{y}$), a large spin-valve signal ($\sim$ 100 m$\Omega$) can be seen even at 150 K. This spin-valve feature is attributed to the change in the magnetization direction of two different Co$_{2}$FeSi contacts used here between nearly parallel and anti-parallel configurations. Note that the magnitude of the nonlocal signal is almost compatible with that observed at 4 K for a device (Device B in Ref. \cite{Zhou}) with Fe/MgO tunnel contacts and $L =$ 1.0 $\mu$m in the previous work reported by Zhou {\it et al}.\cite{Zhou} This feature implies that, even at a higher temperature, we demonstrate the spin accumulation in the {\it n}-Ge channel equivalent to that by Zhou {\it et al}.\cite{Zhou} Using this nonlocal four-terminal geometry, we also applied out-of-plane magnetic field ($B_{Z}$) under parallel and anti-parallel magnetic configurations for the Co$_{2}$FeSi electrodes and recorded $\Delta R_{\rm NL}$ as a function of $B_{Z}$. As a result, clear Hanle-type spin precession curves, which are evidence for the generation, manipulation, and detection of pure spin currents in the {\it n}-Ge channel, can be seen in Fig. 2(b). At further higher temperatures, such Hanle-effect curves were observable for several devices. 

By fitting the Hanle-effect curves with the one-dimensional spin drift diffusion model,\cite{Jedema,Lou} the spin lifetime ($\tau_{s}$) in the channel region can roughly be extracted. The model used for our device is, 
\begin{equation}
\Delta R_{\rm NL}(B_{Z})  \propto  \pm	\int_0^{\infty} \frac{1}{\sqrt{4\pi Dt}}\exp\left[-\frac{L^{2}}{4Dt}\right]\cos(\omega_{L}t) \exp(-\frac{t}{\tau_{s}})dt,  
\end{equation}
where $\pm$ is the sign depending on the magnetization configuration (parallel or anti-parallel), $D$ is the diffusion constant of the Ge channel, $\omega_{L} =$ $g\mu_{B}$$B_{Z}$/$\hbar$ is the Lamor frequency, $g$ is the electron $g$-factor ($g =$ 1.56),\cite{Vrijen} $\mu_{B}$ is the Bohr magneton. The representative fitting results are shown in the solid curves in Fig. 2(b). 
At 150 K, the $\tau_{s}$ and $D$ values for the {\it n}-Ge channel used are estimated to be $\sim$ 420 ps and $\sim$ 8.3 cm$^{2}$s$^{-1}$, respectively. The obtained $D$ value is consistent with experimentally estimated value of 8.6 cm$^{2}$s$^{-1}$ at 150 K from Einstein relation, $D$ $=$ $\frac{k_{B}T}{q}$$\mu_{e}$, where $k_{B}$ is Boltzmann's constant. We will comment on $\tau_{s}$ later. 

From the $|\Delta R_{\rm NL}|$ shown in Fig. 2(b), we also estimate the spin generation efficiency of this device with the Co$_{2}$FeSi/{\it n}$^{+}$-Ge contacts. In general, $|\Delta R_{\rm NL}|$ detected by nonlocal four-terminal geometry can be expressed as,  
\begin{equation}
|\Delta R_{\rm NL}| = \frac{P_{\rm gen}P_{\rm det}\rho_{\rm N}\lambda_{\rm N}}{S} \exp(-\frac{L}{\lambda_{\rm N}}),
\end{equation}
where $P_{\rm gen}$ and $P_{\rm det}$ are spin polarization which are generated and detected, respectively, at the ferromagnetic electrodes. Thus, ($P_{\rm gen}$ $\times$ $P_{\rm det}$)$^{1/2}$ can roughly be regarded as the spin generation efficiency. $\rho_{\rm N}$ and $\lambda_{\rm N}$ are the resistivity and the spin diffusion length of the nonmagnetic channel. In this study, $\rho_{\rm Ge}$ and $\lambda_{\rm Ge}$ at 150 K are 28.2 m$\Omega$cm and 0.59 $\mu$m, respectively, where $\lambda_{\rm Ge} =$$\sqrt{D \tau_{s}}$. $S$ is the cross section of the Ge channel ($\sim$ 5.0 $\mu$m$^{2}$) and we used $L =$ 1.0 $\mu$m. Using these parameters, we can obtain ($P_{\rm gen}$ $\times$ $P_{\rm det}$)$^{1/2}$ of $\sim$ 0.12 ($\sim$ 12 \%). Despite a higher temperature of 150 K the spin generation efficiency is about two orders of magnitude larger than that for an Fe/MgO/Ge device (Device B at 4 K in Ref. \cite{Zhou}). It should be noted that the order of this efficiency is consistent with that for a GaAs based device with $L$2$_\text{1}$-ordered Co$_{2}$FeSi electrodes ($\sim$ 0.16).\cite{Bruski} Even for Ge, the $L$2$_\text{1}$-ordered Co$_{2}$FeSi electrodes enable us to markedly enhance the generation of spin accumulation in the channel. Considering this study for Ge and a previous one for GaAs,\cite{Bruski} the Heusler-compound electrodes can open a possible way for highly efficient spin generation even in semiconductor channels.\cite{Ramsteiner,Bruski,Tezuka1,HamayaJJAP,Kasahara2,Hamaya2}
\begin{figure}
\begin{center}
\includegraphics[width=8.5cm]{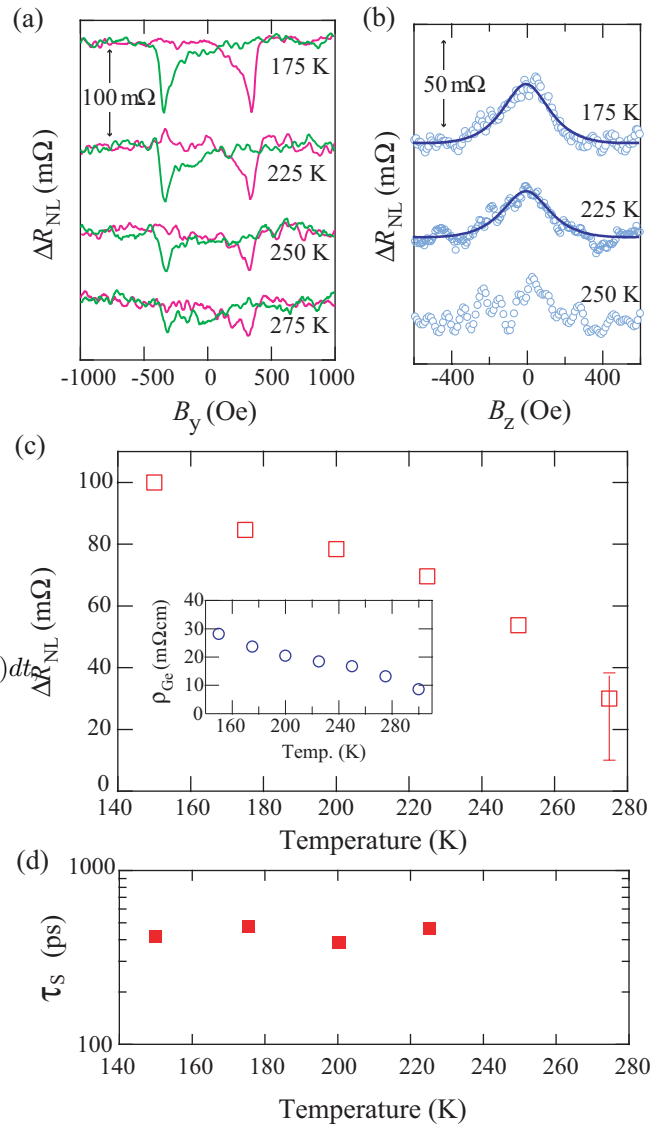}
\caption{(Color online) Temperature dependent (a) NLSV signals, (b) Hanle curves, (c) $\Delta R_{\rm NL}$, and (d) $\tau_{s}$. The inset of (c) shows $\rho_{\rm Ge}$ as a function of temperature for the used channel.}
\end{center}
\end{figure}

Next, we discuss spin signals with temperature evolution for this device. Since the temperature dependent $RA$ can affect the spin signals, shown in our previous work,\cite{Hamaya2} we simply focus on the data which can be measured in the condition of a constant $RA$ ($\sim$ 10$^{4}$ $\Omega$$\mu$m$^{2}$) above 150 K. As shown in Figs. 3(a) and 3(b), nonlocal spin signals and Hanle curves in the parallel magnetic configuration of the Co$_{2}$FeSi electrodes are gradually decreased with increasing temperature. In this experiment, the clear Hanle curves can be observed up to 225 K although spin-valve like nonlocal signal can be seen at 275 K. Unfortunately, both features could not be verified at room temperature. Figure 3(c) shows $\Delta R_{\rm NL}$, which is roughly extracted from NLSV signals, as a function of temperature. In order to understand $\Delta R_{\rm NL}$ versus temperature, we also estimated $\tau_{s}$ from the fitting of the Hanle curves with increasing temperature, where $D$ values were estimated from Einstein relation and experimental transport measurements of carriers in micro Hall-bar devices. Interestingly, Fig. 3(d) reveals that there is no clear dependence with temperature evolution, which is largely different from the features in high-temperature regime in the previous works.\cite{Zhou,Chang} As a whole, the $\tau_{s}$ values ($\sim$ 380 ps $\leqslant$ $\tau_{s}$ $\leqslant$ $\sim$ 480 ps ) extracted here are markedly shorter than theoretically expected intrinsic ones ($\tau_{s}$ $\sim$ 10 ns).\cite{Li} These features of $\tau_{s}$ with temperature evolution cannot be explained by the Elliot-Yafet spin relaxation mechanism.\cite{Zhou,Elliott,Yafet,Guite,Pezzoli} We should also recognize the large influence of the spin-flip scattering due to the heavily doped Sb near the Co$_{2}$FeSi/{\it n}$^{+}$-Ge interface on the spin generation and detection,\cite{Chang} giving rise to the much shorter $\tau_{s}$ values. Taking this special situation for our devices into account, we infer that the $\tau_{s}$ value with temperature evolution is limited by the extrinsic factors rather than conventional spin relaxation mechanism.\cite{Hamaya2} 

According to Eq. (2), we find that the magnitude of $\Delta R_{\rm NL}$ depends not only on $\lambda_{\rm N} (=$$\sqrt{D \tau_{s}}$) but also on $\rho_{\rm N}$. Thus, we should reconsider the change in $\rho_{\rm Ge}$ with temperature evolution. The inset of Fig. 3(c) shows the temperature dependence of $\rho_{\rm Ge}$ for the {\it n}-Ge channel used in this study. Since the electron density of the used Ge channel is reduced with decreasing temperature, $\rho_{\rm Ge}$ decreases monotonically. As expected, this feature is close to the temperature dependent $\Delta R_{\rm NL}$ in the main panel of Fig. 3(c). From these considerations, we can judge that the contribution of $\rho_{\rm Ge}$ to $\Delta R_{\rm NL}$ is more dominant than that of the spin relaxation in the Ge channel in this study. 

We finally comment on the short $\lambda_{\rm Ge}$ of $\sim$ 0.6 $\mu$m at low temperatures. As described before, our Ge channels were fabricated by a conventional ion implantation technique and then, we used some etching processes to fabricate devices for transport measurements. Thus, a lack of optimization of these processes gave rise to damaged channels with relatively poor mobility ($\sim$ 300 cm$^{2}$V$^{-1}$s$^{-1}$ at 300 K) compared to bulk wafers (more than $\sim$ 1000 cm$^{2}$V$^{-1}$s$^{-1}$).\cite{Nakaharai,Chui,Satta} As a result, the $D$ values of $\sim$10 cm$^{2}$s$^{-1}$ are relatively low at low temperatures, leading to short $\lambda_{\rm Ge}$. Even though we detect the lateral transport of pure spin currents in {\it n}-Ge at higher temperatures by using Heusler-compound Co$_{2}$FeSi electrodes because of the large enhancement in the spin generation efficiency, a marked improvement of $\lambda_{\rm Ge}$ might also be required for achieving Ge spintronics. Now, optimization of the fabrication processes for lateral structures is underway for obtaining long $\lambda_{\rm Ge}$ in Ge channels for actual device applications.

%One possible origin is the influence of the spin-flip scattering due to the influence of the heavily doped Sb near the Co$_{2}$FeSi/{\it n}$^{+}$-Ge interface on the spin generation and detection.\cite{Chang} Another one is the influence of the position of the Fermi energy ($E_{F}$) in the half-metallic gap for Co$_{2}$FeSi. In general, the $E_{F}$ position of Co$_{2}$FeSi is located near the bandedge of the minority spin band, giving rise to the strong reduction in the spin polarization at higher temperatures.\cite{Tezuka1,Bruski,Inomata} If the influence of the $E_{F}$ position of the Co$_{2}$FeSi electrodes on the spin generation efficiency is large, the spin accumulation in Ge, generated by the Co$_{2}$FeSi/{\it n}$^{+}$-Ge contacts, is markedly reduced at higher temperatures. Although the precise mechanism is unclear yet, we will further explore other Heusler-compound electrodes such as Co$_{2}$FeSi$_{0.5}$Al$_{0.5}$ and Co$_{2}$FeAl because of the change in the $E_{F}$ position from now on. 

In summary, we showed large enhancement in spin generation efficiency in {\it n}-Ge by using $L$2$_\text{1}$-ordered Co$_{2}$FeSi Heusler-compound electrodes. Thanks to this technological development, we demonstrated the lateral transport of pure spin currents in {\it n}-Ge up to 225 K. For actual Ge spintronic applications, the improvement of the device fabrication processes is also required. 

%\vspace{3mm}
%\begin{acknowledgments} 
This work was partly supported by Industrial Technology Research Grant Program from NEDO and  Grant-in-Aid for Scientific Research (A) (No. 25246020) from The Japan Society for the Promotion of Science (JSPS). S.Y. acknowledges JSPS Research Fellowships for Young Scientists.
% \end{acknowledgments} 

%\clearpage
%Create the reference section using BibTeX:
%\noindent{{\bf References}}

%\clearpage
%\noindent{{\bf Figure Captions}}
%\vspace{8mm}

%{\small\noindent Fig. 1: {(Color online) (a) Schematic diagram of a lateral four-terminal devices with Co$_{2}$FeSi/{\it n}$^{+}$-Ge contacts. (b) A cross-sectional TEM image of the Co$_{2}$FeSi/{\it n}$^{+}$-Ge interface fabricated by a room-temperature MBE technique. (c) Nanobeam electron diffraction patterns for each Co$_{2}$FeSi and {\it n}$^{+}$-Ge. The zone axis is parallel to the [011] direction.} 
%\vspace{8mm}

%{\small\noindent Fig. 2: {(Color online) (a) Nonlocal magnetoresistance curve at 150 K. (b) Hanle-effect curves at 150 K for paralel and anti-parallel magnetization configuration. The $|\Delta R_{\rm NL}|$ value is defined from the Hanle curves.} 
%\vspace{8mm}

%{\small\noindent Fig. 3: {(Color online)Temperature dependent (a) NLSV signals, (b) Hanle curves, (c) $\Delta R_{\rm NL}$, and (d) $\tau_{s}$. The inset of (c) shows $\rho_{\rm Ge}$ as a function of temperature for the used channel..} 
%\vspace{8mm}

%\clearpage
%\begin{figure}
%\vspace{20mm}
%\begin{center}
%\includegraphics[width=8.5cm]{Fig1.eps}
%\caption{(Color online) Schematic diagrams of lateral (a) three-terminal and (b) four-terminal devices with high-quality Fe$_{3}$Si/{\it n}$^{+}$-Ge contacts. (c) Temperature dependent $I_{21}$-$V_{21}$ characteristics measured for the three-terminal device. (d) Enlarged scanning electron micrograph of the contacts for the four-terminal device. }
%\end{center}
%\end{figure}

\end{document}